\documentclass[twocolumn]{aastex62}
\usepackage{graphicx}	
\usepackage{amsmath}	
\usepackage{amssymb}	
\usepackage{verbatim}
\usepackage{color}
\received{January 1, 2018}
\revised{January 7, 2018}
\accepted{January 7, 2018}
\submitjournal{ApJ}

\shorttitle{Thermal Tracer of Icy Mantles}
\shortauthors{He et al.}

\begin{document}

\title{The $^{12}$CO$_2$ and $^{13}$CO$_2$ Absorption Bands as Tracers of the Thermal History of Interstellar Icy Grain
 Mantles}

\correspondingauthor{Jiao He}
\email{jhe08@syr.edu}

\correspondingauthor{Gianfranco Vidali}
\email{gvidali@syr.edu}

\author[0000-0003-2382-083X]{Jiao He}
\affiliation{Physics Department, Syracuse University, Syracuse, NY 13244, USA}
\affiliation{Current address: Raymond and Beverly Sackler Laboratory for Astrophysics, Leiden Observatory, Leiden University, PO Box 9513, 2300 RA Leiden,The Netherlands}

\author{SM Emtiaz}
\affiliation{Physics Department, Syracuse University, Syracuse, NY 13244, USA}

\author[0000-0001-9344-0096]{Adwin Boogert}
\affiliation{Institute for Astronomy, University of Hawai'i at Manoa, 2680 Woodlawn Drive, Honolulu, HI 96822–1839, USA}

\author[0000-0002-4588-1417]{Gianfranco Vidali}
\affiliation{Physics Department, Syracuse University, Syracuse, NY 13244, USA}

\begin{abstract}
Analyses of infrared signatures of CO$_2$ in water dominated ices in the ISM
can give information on the physical state of CO$_2$ in icy grains and on the
thermal history of the ices themselves. In many sources, CO$_2$ was found in
the ``pure'' crystalline form, as signatured by the splitting in the bending
mode absorption profile. To a large extent, pure CO$_2$ is likely to have formed from
segregation of CO$_2$ from a CO$_2$:H$_2$O mixture during thermal processing.
Previous laboratory studies quantified the temperature dependence of
segregation, but no systematic measurement of the concentration dependence of
segregation is available. In this study, we measured both the temperature
dependence and concentration dependence of CO$_2$ segregation in CO$_2$:H$_2$O
mixtures, and found that no pure crystalline CO$_2$ forms if the CO$_2$:H$_2$O
ratio is less than 23\%. Therefore the segregation of CO$_2$ is not always
a good thermal tracer of the ice mantle. We found that the position and width
of the broad component of the asymmetric stretching vibrational mode of $^{13}$CO$_2$ change linearly with the temperature of CO$_2$:H$_2$O mixtures, but are insensitive to the
concentration of CO$_2$. We recommend using this mode, which will be observable towards low mass protostellar envelopes
and dense clouds with the James Webb Space Telescope, to trace the thermal
history of the ice mantle, especially when segregated CO$_2$ is unavailable. We
used the laboratory measured $^{13}$CO$_2$ profile to analyze the ISO-SWS
observations of ice mantles towards Young Stellar Objects, and the
astrophysical implications are discussed.
\end{abstract}


\section{Introduction}
CO$_2$ is abundant in quiescent and star-forming molecular clouds where it is
found in ices with abundance in the 10 to 50\% range with respect to water.
Solid state CO$_2$ is mostly detected in the mid-infrared absorption through
the asymmetric stretching mode $\nu_3$ at $\sim$2350 cm$^{-1}$
\citep{Gerakines1999, Nummelin2001,Noble2013} and through the bending mode at
$\sim$665~cm$^{-1}$ \citep{Gerakines1999, Pontoppidan2008, Ioppolo2013,
Noble2013}. Additional modes are detected as well, such as the combination
modes $\nu_1+\nu_3$ at 3708 cm$^{-1}$ and $2\nu_2+\nu_3$ at 3600 cm$^{-1}$
\citep{Gerakines1999, Keane2001}. Unlike CO, which only has a high abundance in
highly shielded regions, CO$_2$ has the same threshold of formation as water
\citep{Bergin2005, Whittet2009}, which means that CO$_2$ is mixed with water in
pristine polar ices coating dust grains \citep{Whittet2009}. The column density
ratio of CO$_2$:H$_2$O varies between 10\% and 50\%, depending on the specific
cloud. Laboratory measurements of CO$_2$-containing ice mixtures have found
that the infrared absorption profile of CO$_2$ strongly depends on the physical
and chemical environment in the ice, such as the temperature and the ice
compositions \citep{Gerakines1999, Oberg2009, Hodyss2008, He2018}. This makes
CO$_2$ a very good candidate to trace the composition and physical condition of
the ice mantle.

Since the $\nu_3$ asymmetric stretch is particularly strong and often
saturated, the $\nu_2$ bending mode at $\sim650$ cm$^{-1}$ and the
$^{13}$CO$_2$ asymmetric stretching mode at $\sim$2280 cm$^{-1}$ are often used
instead to study CO$_2$ in  ice mantles. So far, most of the observations of
solid state CO$_2$ are through the 650 cm$^{-1}$ feature.
\citet{Pontoppidan2008} used Spitzer to systematically study the 650 cm$^{-1}$
feature in Young Stellar Objects (YSOs). By comparing the observed absorption profile with laboratory
measurements of different CO$_2$-containing ice mixtures, they found that the
observed spectra can be fit well using five different components, each
representative of CO$_2$ in different ice mixtures measured in the laboratory.
However, there are redundancies in the derived ice compositions that are
further amplified by the effect of the poorly constrained grain shapes on the
observed spectral profiles. The latter is not the case for
$^{13}$CO$_2$ because it is diluted by almost two orders of magnitude in the
ice, resulting in a low polarizability and thus negligible grain shape effects
\citep{Boogert2000}. Therefore, the observed $^{13}$CO$_2$ spectra can
be readily compared with laboratory measured spectra. Its asymmetric stretching
feature cannot be observed with ground-based telescopes because of strong
telluric absorption. The only comprehensive study of this feature is by
\citet{Boogert2000} who used ISO-SWS to observe $^{13}$CO$_2$ in 13 sightlines. However, due to limited sensitivity, only a small sample of sightlines
could be observed, lacking in particular the envelopes of low mass YSOs and
quiescent dense molecular clouds. The forthcoming James Webb Space Telescope
(JWST) is expected to cover this spectral region at orders of magnitude better
sensitivity and somewhat higher spectral resolution
($R=\lambda/\Delta\lambda\sim$3000 versus 2000). A comprehensive set of  laboratory
measurements of solid state $^{13}$CO$_2$ would facilitate the interpretation
of JWST observations of solid CO$_2$. One of the motivations of this work is to
measure the absorption profile of $^{13}$CO$_2$ $\nu_3$ mode in CO$_2$:H$_2$O
mixtures at different mixing ratios and different temperatures, providing
improved insights into the composition as well the thermal history of
interstellar and circumstellar icy mantles.

In some of the sightlines observed by \citet{Pontoppidan2008} and
\citet{Gerakines1999}, the CO$_2$ bending profile shows double splitting
features. This is interpreted as the Davydov splitting---which occurs in
crystals with more than one identical molecular species or unit per unit cell.
It is commonly interpreted as an indication of ``pure'' crystalline CO$_2$
\citep{Pontoppidan2008, isokoski2013, Cooke2016, Baratta2017}. Since ices in
the ISM are water-dominated, the appearance of the splitting indicates the
formation of segregated CO$_2$ solids due to thermal processing.
\citet{Oberg2009} measured the temperature dependence of the segregation of
CO$_2$ from CO$_2$:H$_2$O mixtures. The majority of their experiments were
carried out with a CO$_2$:H$_2$O mixing ratio of 1:2, with only one measurement
for 1:4, a more representative ratio for the ices coating grains
\citep{Boogert2015}. From an Arrhenius fitting to the experimental data, they
found an energy barrier of 1090$\pm$15~K for segregation of CO$_2$. This
translates in a segregation temperature of 30$\pm$5~K, assuming a segregation
time scale of 4000 yrs. \citet{He2017} obtained an onset of segregation of
CO$_2$ on the surface of non-porous Amorphous Solid Water (np-ASW) at 65~K,
corresponding to a temperature in space of 43$\pm$3~K, assuming a diffusion
pre-exponential factor of $10^{12}$ s$^{-1}$ and a similar segregation time
scale as in \citet{Oberg2009}. This temperature range is somewhat higher than
the result of \citet{Oberg2009}.

Since CO$_2$ is present in a wide range of concentrations with respect to water
in ices \citep{Boogert2015,Yamagishi2015}, in order to correctly interpret IR
spectra for studying the thermal evolution of ices, it is necessary to know how
the level of CO$_2$ concentration in mixed CO$_2$:H$_2$O ices impacts
segregation. In this work, we comprehensively study the concentration
dependence as well as temperature dependence of CO$_2$ segregation from
CO$_2$:H$_2$O mixtures. In a previous work \citep{He2018}, we found that the
$\nu_1+\nu_3$ mode at $\sim3708$ cm$^{-1}$ and the $\nu_1+2\nu_2$ mode at
$\sim3600$ cm$^{-1}$ provide useful tools to quantify CO$_2$ segregation in
laboratory measurements. In this work, we study the profiles of the combination
modes and show that they can be used to assess the degree of order of CO$_2$ in
CO$_2$:H$_2$O ices. To obtain the thermal history of a CO$_2$:H$_2$O ice and
to compare it with spectra obtained with the ISO-SWS, we use the $\nu_3$
absorption profile of $^{13}$CO$_2$ naturally occurring in laboratory
CO$_2$:H$_2$O ice mixtures.

The remaining of this paper is organized as follows: Section~\ref{sec:exp}
describes the experimental setup, followed by Section~\ref{sec:results} on results and
analysis. Section~\ref{sec:astro} compares our laboratory measured spectra of
$^{13}$CO$_2$ with ISO-SWS data and discusses the astrophysical implications.

\section{Experimental setup}
\label{sec:exp}
Experiments were carried out in a ultra-high vacuum (UHV) apparatus at
Syracuse University. The UHV chamber is pumped by a combination of turbopumps
and a cryopump. After bake-out, the base pressure reaches $4\times10^{-10}$
Torr routinely. At the center of the UHV chamber, a gold-coated copper disc was
used as the sample. The sample can be cooled down to 5~K by an Advanced
Research Systems DE-204S cryocooler, or heated up to room temperature using
a cartridge heater located right behind the sample. The sample temperature was
measured by a calibrated silicone diode to an accuracy better than 50 mK.
A Lakeshore 336 temperature controller was used to read and control the
temperature.

The IR spectra of ices deposited on the sample were recorded using a Nicolet
6700 Fourier Transform InfraRed (FTIR) in the Reflection Absorption InfraRed
Spectroscopy (RAIRS) configuration with an incident angle of $\sim78$ degrees.
The FTIR collects and averages 9 spectra every 20 seconds at a resolution of
0.5 cm$^{-1}$ in the range of 600--4000 cm$^{-1}$. Because of the strong
signals, we took averages of fewer scans than it is typically done in order to
obtain a good time resolution during warming up of the ice sample. The heating ramp
rate during temperature programmed desorption (TPD) was 0.1 K/s (except for
dedicated flash heatings), which amounts to one infrared spectrum every 2 K.
The modalities of deposition of CO$_2$:H$_2$O mixtures onto the gold-plated copper sample are discussed in the Appendix.

\section{Results and Analysis}
\label{sec:results}
We carried out three sets of experiments. In the first set, we study the
temperature dependence of IR absorption bands of CO$_2$:H$_2$O mixtures of
different mixing ratios as they were heated linearly from 10 K to 200 K. In the
second set, we fix the CO$_2$:H$_2$O ratio and carried out isothermal experiments
at different temperatures to find out the temperature at which CO$_2$
segregation maximizes. A higher temperature facilitates segregation, but at
too high a temperature, CO$_2$ desorption begins to compete with segregation.
There should exist an optimum temperature that maximizes segregation. In the
third set of experiments, we fix the temperature for the isothermal experiments
at the temperature of maximum segregation we found from the second set of
experiments, and check how segregation depends on CO$_2$ concentration.

\subsection{Temperature dependence of IR bands}\label{sec:linear}
\begin{figure}
  \includegraphics[width=1\columnwidth]{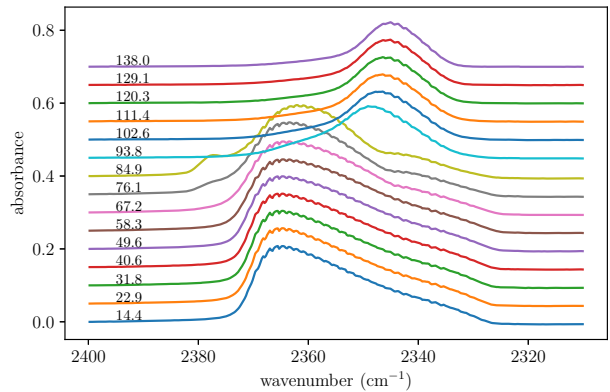}
  \caption{RAIRS of $\nu_3$ band of 50:100 CO$_2$:H$_2$O mixture deposited at
  10~K and heated at 0.1~K/s. The temperature of each curve is marked. The
  small oscillations superimposed on the curves are due to gas-phase CO$_2$ in
  the spectrometer.} \label{fig:2350}
\end{figure}

\begin{figure}
  \includegraphics[width=1\columnwidth]{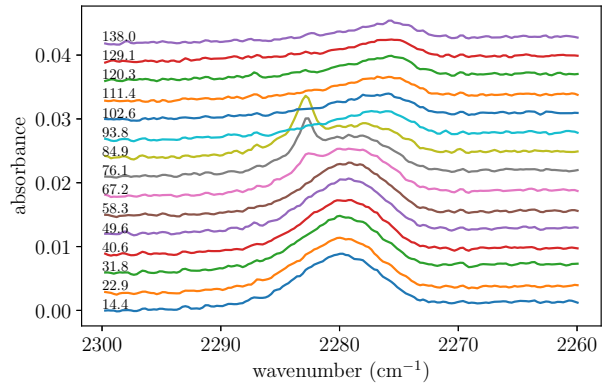}
  \caption{Same as Figure~\ref{fig:2350},but for the $\nu_3$ band of $^{13}$CO$_2$ that is present in natural abundance in CO$_2$:H$_2$O ice
mixtures.}\label{fig:2280}
\end{figure}

\begin{figure}
  \includegraphics[width=1\columnwidth]{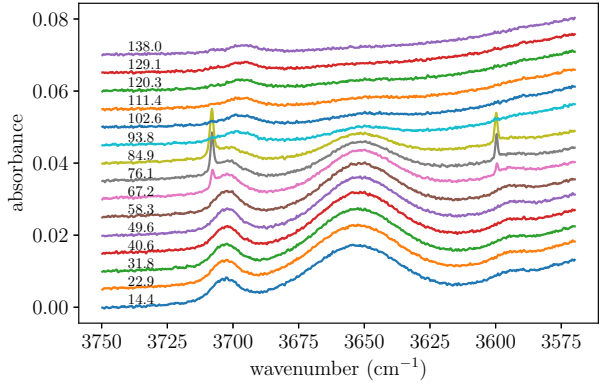}
  \caption{Same as Figure~\ref{fig:2350}, but for the 3570 cm$^{-1}$ to 3750
  cm$^{-1}$ range.}\label{fig:3600}
\end{figure}

\begin{figure}
  \includegraphics[width=1\columnwidth]{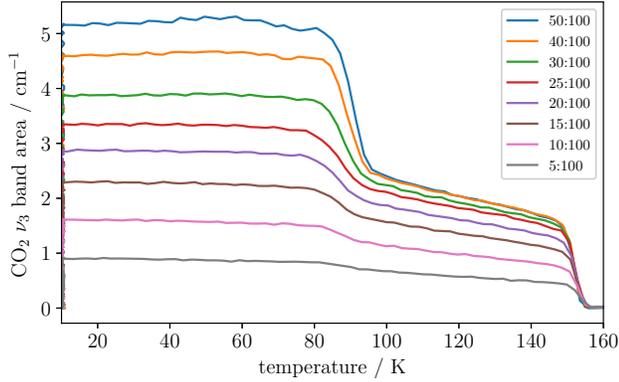}
  \caption{Band area of the $\nu_3$ absorption peak of CO$_2$ during warming up
  of the CO$_2$:H$_2$O mixtures with different mixing ratios (see
  inset).}\label{fig:trapping}
\end{figure}

In this set of experiments, 50 ML of water and various amount of CO$_2$ were
co-deposited on the sample at 10 K, to make the following CO$_2$:H$_2$O
mixtures: 5:100, 10:100, 15:100, 20:100, 25:100, 30:100, 40:100, and 50:100.
After deposition, the ice mixtures were heated linearly from 10 K to 200 K at
0.1~K/s. Figures~\ref{fig:2350}, \ref{fig:2280}, and
\ref{fig:3600} show the absorption bands $\nu_3$ , $^{13}$CO$_2$ $\nu_3$, and
the combination modes of a 50:100 CO$_2$:H$_2$O mixture at selected temperatures
during heating. Figure~\ref{fig:trapping} shows the integrated band area of
the $\nu_3$ peak at around 2350 cm$^{-1}$ for CO$_2$:H$_2$O mixtures of
different mixing ratios during the heating up.

The CO$_2$ $\nu_3$ mode shows an asymmetric peak centered at about 2365
cm$^{-1}$ at low temperatures. \citet{He2018} has shown that the $\nu_3$ peak
of CO$_2$ on water ice surface is dependent on the CO$_2$ coverage. As the
coverage increases from zero to more than 2 layers (L), the peak shifts from
$\sim2347$ cm$^{-1}$ to $\sim2376$ cm$^{-1}$. The shape and position of the
spectra in Figure~\ref{fig:2350} at low temperatures are qualitatively similar
to the submonolayer spectra shown in Figure 1 of \citet{He2017}. As the ice is
heated to the 70--80 K range, a peak at $\sim 2378$ cm$^{-1}$ emerges, which is a signature of ``pure'' crystalline form of CO$_2$ after segregation
has taken place. Between 80 and 100 K, the $\nu_3$ band area of CO$_2$
decreases. This is due to the desorption of weakly bound CO$_2$ on the surface
of water ice (including the surface of pores). The remaining CO$_2$ molecules that are
trapped inside the water ice matrix have an absorption peak at around $2345$
cm$^{-1}$ that redshifts with temperature. Between 100 K and 150 K, the CO$_2$
amount decreases linearly with temperature. This slow desorption is induced by the
compaction of ASW, and CO$_2$ molecules are pushed out of the water matrix
during the pore collapse of ASW. Between 150 K and 155 K, water crystallizes
and all of the remaining CO$_2$ desorb from the ice. This is referred to as
the ``molecular volcano desorption'' \citep{Smith1997}.

\begin{figure}
  \includegraphics[width=1\columnwidth]{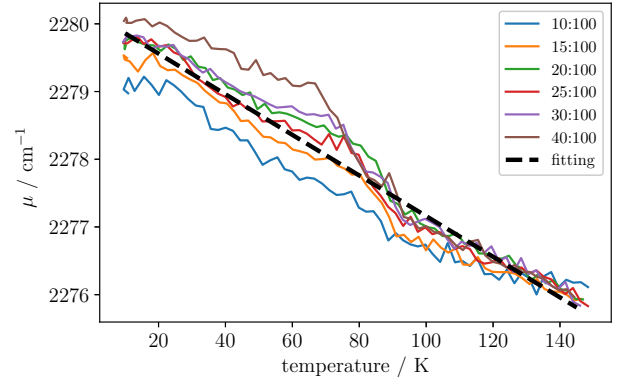}
  \caption{Center position ($\mu$) of the Gaussian disordered component of the
  $^{13}$CO$_2$ $\nu_3$ absorption peak at $\sim$2280 cm$^{-1}$ for different
  ratios of CO$_2$:H$_2$O mixtures during warm-up. The CO$_2$:H$_2$O ratio is
  in the inset.}\label{fig:mu}
\end{figure}

\begin{figure}
  \includegraphics[width=1\columnwidth]{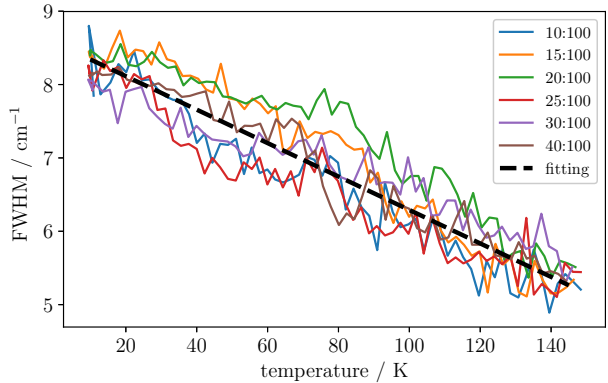}
  \caption{Same as in Figure~\ref{fig:mu} but for $FWHM$.}\label{fig:sigma}
\end{figure}

A similar trend is also seen in the $\nu_3$ $^{13}$CO$_2$ band. When CO$_2$
segregates, a peak at 2282 cm$^{-1}$ emerges. This peak is characteristic of $^{13}$CO$_2$ with natural isotopic abundance
\citep{He2017}, and is more sensitive to segregation than the $\nu_3$ peak of
CO$_2$ at around 2376 cm$^{-1}$. During the heating from 10 K to 140 K, the
peak red shifts from about 2280 cm$^{-1}$ to about 2276 cm$^{-1}$, and the
width becomes narrower with temperature. To see more clearly how the peak
position and width change with temperature, we used one broad Gaussian
lineshape and one narrow Lorentzian lineshape to fit $^{13}$CO$_2$ in
disordered and ordered (crystalline) CO$_2$, respectively. Although the
disordered component has an asymmetric shape, for simplicity we still use
a Gaussian function. This fitting scheme is sufficient to reliably obtain the
peak position and the width. Figure~\ref{fig:mu} and~\ref{fig:sigma} show the
center position ($\mu$) and the $FWHM$ of the Gaussian fit, respectively. Both
the center position $\mu$ and $FWHM$ decrease with temperature roughly
linearly. The best fitting parameters with a 95\% confidence interval are
\begin{align}
  &\mu=(2280.16\pm0.06) - (0.030\pm0.0006)T \label{eq:mu} \\
&FWHM=(8.6\pm0.7) - (0.022\pm0.001)T \label{eq:sigma}
\end{align}
where the unit is cm$^{-1}$ for $\mu$ and $FWHM$, and Kelvin for $T$. These
simple functions will be useful for the analysis of the observed $^{13}$CO$_2$
profile, to be discussed below.

Figure~\ref{fig:3600} shows the region of the $\nu_1+\nu_3$ $^{12}$CO$_2$ mode
at around 3700 cm$^{-1}$, $\nu_1+2\nu_2$ $^{12}$CO$_2$ mode at 3600 cm$^{-1}$,
as well as a peak at $\sim$3650 cm$^{-1}$. The 3650 cm$^{-1}$ peak can be
attributed to $^{12}$CO$_2$ on the surface of water. This feature is common in
CO$_2$:H$_2$O mixtures or CO$_2$ on the surface of water at low temperatures.
The $\nu_1+\nu_3$ and $\nu_1+2\nu_2$ combination modes are broad at low
temperatures. As the CO$_2$ segregates, a sharp feature emerges for both
combination modes. After weakly bound CO$_2$ has desorbed from the surface, the
3600 cm$^{-1}$ peak is barely seen, but the peak at 3700
cm$^{-1}$ is still clearly visible. Its position and width are no different
from the 3-coordinated dangling bond of amorphous water annealed at similar
temperatures. We attribute this peak to the dangling bond of ASW, although we
do not exclude the possibility that the $\nu_1+\nu_3$ mode of CO$_2$ may also
have a small contribution to this peak.

\subsection{Segregation of CO$_2$ in CO$_2$:H$_2$O mixtures} \label{sec:segregation}
Prior laboratory measurements \citep{Hodyss2008,He2018} have shown that the
segregation and crystallization of CO$_2$ is accompanied by changes in the
bending, asymmetric stretching (for both $^{12}$CO$_2$ and $^{13}$CO$_2$), and
combination modes. The bending mode absorption at $\sim$650 cm$^{-1}$ is an
important feature that is used to characterize CO$_2$ ice
\citep{Pontoppidan2008}. Toward heavily embedded YSOs, the bending mode is
easier to observe than the combination modes or the $\nu_3$ of $^{13}$CO$_2$
because of the brighter continuum emission at 15 $\mu$m. However, the bending
mode band is close to the lower limit of our infrared detector, and the signal
is weak. Following our previous work on CO$_2$ ice \citep{He2018}, we use the
$\nu_1+\nu_3$ combination mode at around 3700 cm$^{-1}$ to quantify the
segregation. The analysis of segregation based on the combination mode at 3700
cm$^{-1}$should not differ from the that using the bending mode at $\sim$650
cm$^{-1}$.

\begin{figure}
  \includegraphics[width=1\columnwidth]{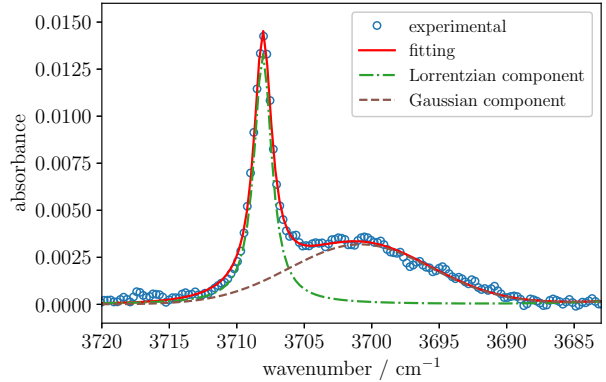}
  \caption{An example of fitting the CO$_2$ $\nu_1+\nu_3$ mode absorption profile using a Lorentzian component and a Gaussian component.}\label{fig:fit}
\end{figure}

\begin{figure}
  \includegraphics[width=1\columnwidth]{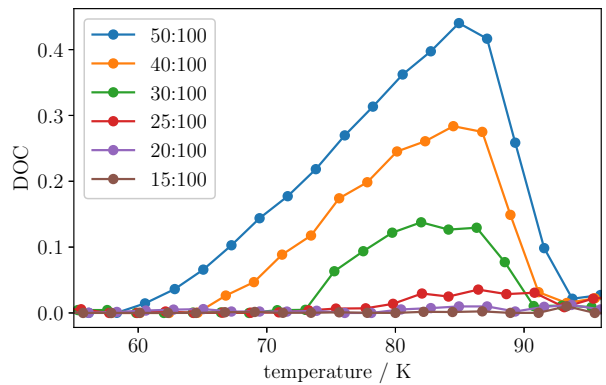}
  \caption{The degree of crystallinity ($DOC$) of CO$_2$:H$_2$O mixtures when warming up the ice at a ramp rate of 0.1~K/s. The mixing ratios are shown in the figure.} \label{fig:linear_doc}
\end{figure}

We decompose the profile of the combination mode at $\sim$3700 cm$^{-1}$ into
two components, one broad Gaussian component centered at $\sim$3703 cm$^{-1}$
attributed to disordered CO$_2$, and one sharp Lorentzian component centered at
3708 cm$^{-1}$ attributed to (poly)crystalline CO$_2$. An example of the
fitting is shown in Figure~\ref{fig:fit}. We defined the ``degree of
crystallinity'' ($DOC$) as the fraction of CO$_2$ in the (poly)crystalline form
(the Lorentzian component).
\begin{equation}
DOC = \frac{A_{crystalline}}{A_{crystalline}+A_{amorphous}}
\label{eq:doc}
\end{equation}
where $A_{crystalline}$ and $A_{amorphous}$ are the band area of the Lorentzian
component and Gaussian component, respectively. We calculate $DOC$ of the
experiments presented in Section~\ref{sec:linear}. The results are shown in
Figure~\ref{fig:linear_doc}.

\begin{figure}
  \includegraphics[width=1\columnwidth]{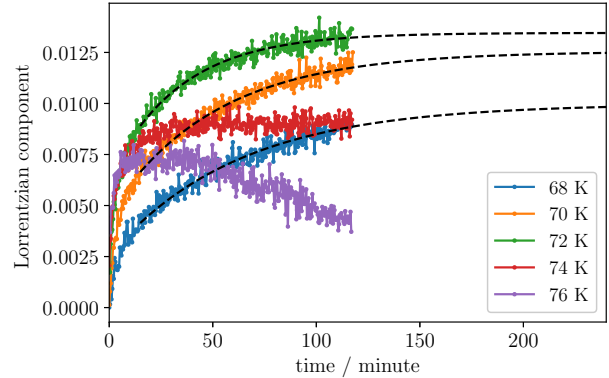}
  \caption{The band area of the Lorentzian component of $\nu_1+\nu_3$ at 3708 cm$^{-1}$ during isothermal experiment of a 30:100 CO$_2$:H$_2$O mixture at different temperatures. The isothermal temperature is marked in the figure. The dashed lines are the fitting using a function $a(1-exp(-bt))+c$. } \label{fig:annT}
\end{figure}

From Figure~\ref{fig:linear_doc}, it can be seen that the segregation strongly
depends on both concentration and temperature. Here we did the experiments at
a fixed heating ramp rate. It is possible that different ramp rates would also
yield different segregation ratios. To use CO$_2$ segregation to trace the
temperature history of the ice mantle would require a systematically study of the
segregation over the whole parameter space of time, temperature, and
concentration, which is a very time consuming set of measurements. Here we take a simpler approach
and focus on one parameter at a time. A set of isothermal experiments were
devoted to find out the temperature that maximizes the segregation. We
deposited 50 ML of water and 30 ML of CO$_2$ simultaneously onto the sample at
10 K, then flash heated the sample at a rate of 0.5 K/s to a target
temperature and kept it at that temperature for 2 hours while monitoring the
$\nu_1+\nu_3$ mode. The band area of the Lorentzian component at different
target temperatures as a function of time is shown in Figure~\ref{fig:annT}.
It can be seen that 72 K is the most favorable temperature to form crystalline
CO$_2$ in CO$_2$:H$_2$O mixtures.
Below 72 K the mobility of CO$_2$ is not
enough for segregation of CO$_2$ to occur to the fullest extent, while above 72
K the desorption of CO$_2$ starts to play a significant role. At 70 and 68 K, the segregated CO$_2$ does not reach maximum after 2 hours. To verify that the highest degree of segregation is indeed achieved at 72 K instead of 68 K or 70 K, we use a function to fit the curves and try to find the saturation level.  \citet{Oberg2009} found that the segregation during isothermal experiments cannot be fit by a single exponential function. Two exponential functions are required to fit it. They attributed the two parts of the segregation to two distinct mechanisms of segregation---surface processes and bulk processes. Here we focus on the second part of the segregation only and use the function $a(1-exp(-bt))+c$ to fit the 68--72 K curves after the first 10 minutes. The fitting are extrapolated to 4 hours to show the saturation level, from which it is clearly that 72 K is the favorable temperature that maximizes the segregation. Based on Figure~\ref{fig:linear_doc}, the temperature at
which the $DOC$ maximizes is similar for all concentrations. Therefore, it is
fair to assume that this favorable temperature 72 K works for all
concentrations.

\begin{figure}
  \includegraphics[width=1\columnwidth]{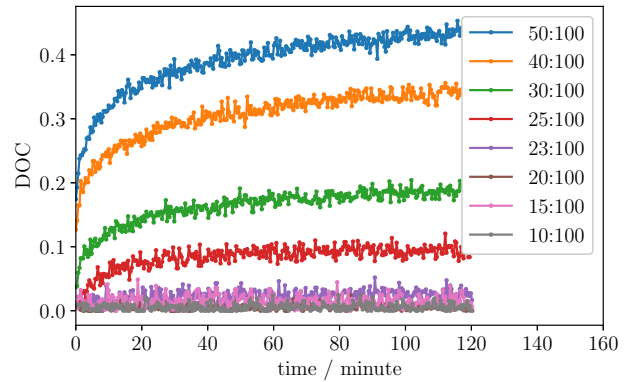}
  \caption{The degree of crystallinity ($DOC$) of CO$_2$:H$_2$O mixture for different concentrations (see inset) during isothermal experiments at 72 K.} \label{fig:ratio}
\end{figure}

After finding this most favorable isothermal experimental temperature, we fix
the temperature and try different concentrations to obtain the lowest
concentration required for the formation of ``pure'' crystalline CO$_2$. We fixed the amount of
water deposited at 50 ML and selected the CO$_2$ dose to be: 2.5, 5.0, 7.5, 10,
11.5, 12.5, 15, 20, and 25 ML. After the co-deposition at 10 K, the ice mixtures
were heated to 72 K at a ramp rate of 0.5 K/s and then kept at 72 K for 2 hours.
Figure~\ref{fig:ratio} shows the $DOC$ as a function of isothermal experimental
time at different target temperatures. It can be seen that below the
CO$_2$:H$_2$O=23:100 concentration, the $DOC$ is almost zero. We thus conclude
that 23\% is the threshold concentration to obtain ``pure'' crystalline CO$_2$
in CO$_2$:H$_2$O mixtures.

\section{Astrophysical Implications}
\label{sec:astro}
\subsection{CO$_2$:H$_2$O ices as Temperature Tracers}
CO$_2$ is one of the main components of ISM ice mantles. In some of the
observed sightlines, CO$_2$ is in the ``pure'' crystalline form, as seen
from the double splitting of the bending absorption profile. This splitting
feature has been proposed to be a candidate tracer of the thermal history of
the ice mantle. Prior laboratory studies \citep{
Ehrenfreund1999,Hodyss2008,Oberg2009} have found that the segregation of CO$_2$ from ice
mixtures is a function of temperature, and the segregation is irreversible with
temperature. Experiments in this work show that the segregation of
CO$_2$ from a CO$_2$:H$_2$O mixture is not only temperature dependent but also
strongly affected by the concentration of CO$_2$. In fact, if the concentration
of CO$_2$ is too low, pure crystalline CO$_2$ never forms, regardless of the
thermal history. According to our measurements, the concentration threshold for
CO$_2$ segregation is 23:100. This is larger than the average CO$_2$/H$_2$O
column density ratio observed toward massive YSOs (17$\pm$3\%;
\citet{Gerakines1999}) and comparable to the $\sim 25$\% ratio of the low mass
YSOs \citep{Pontoppidan2008}. The segregated CO$_2$ detected in these
sightlines (e.g., S140 IRS1) might thus probe CO$_2$/H$_2$O concentrations that
are enhanced at certain locations along the sightline or in certain ice
layers. 

\begin{figure}
 \includegraphics[width=1\columnwidth]{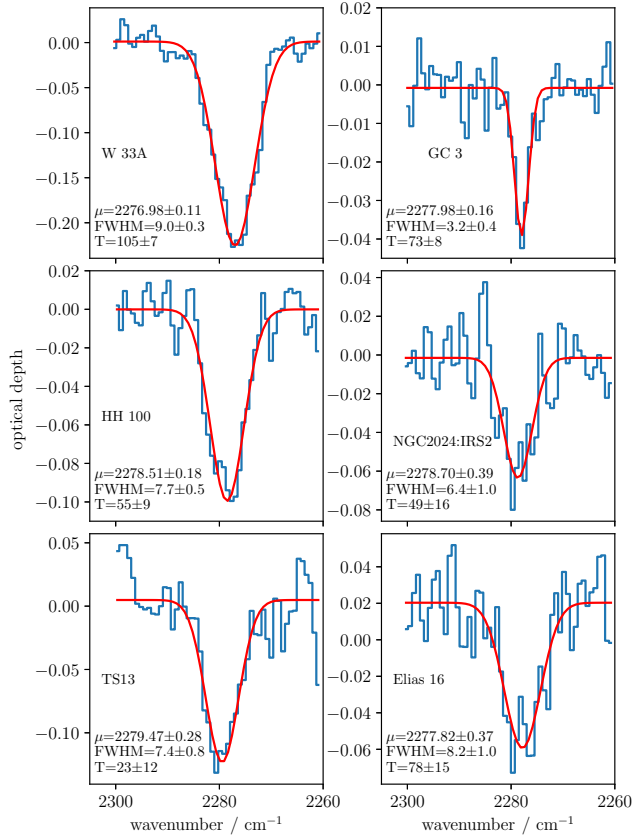}
  \caption{Fitting of setected ISO-SWS spectra of $^{13}$CO$_2$ using one Gaussian function. The fitting parameters are shown in the figure. The temperature $T$ is calculated using Eq~\ref{eq:mu}, and corresponds to the temperature in the laboratory time scale. To convert the time scale to that of the warming up stage of a interstellar clouds, the temperature should be multiplied by a factor of 0.3--0.5.} \label{fig:1comp}
\end{figure}

\begin{figure}
  \includegraphics[width=1\columnwidth]{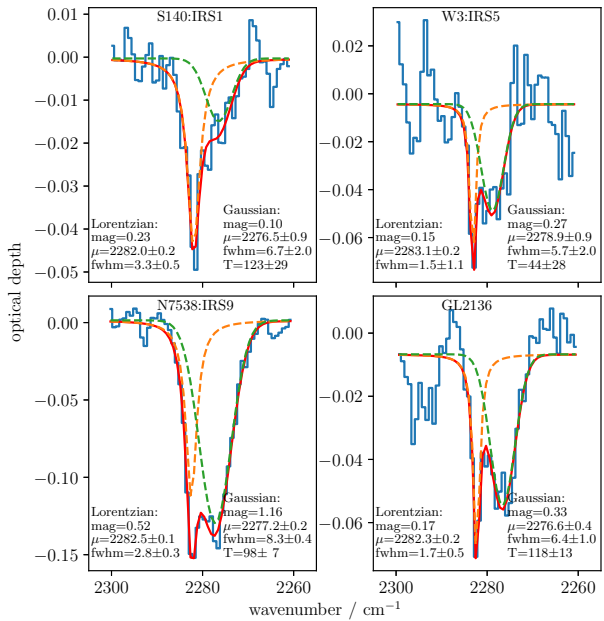}
 \caption{Fitting of selected ISO-SWS spectra of $^{13}$CO$_2$ using one Gaussian distribution and one Lorentzian distribution. The fitting parameters are shown in the figure. The temperature $T$ refers to the laboratory time scale.} \label{fig:2comp}
\end{figure}

Our experiments show that the $^{13}$CO$_2$ absorption profile at around 2280
cm$^{-1}$ is a good tracer of the thermal history of the ice mantle, even if
there is no sign of the very narrow feature of segregated crystalline CO$_2$.
Figure~\ref{fig:mu} and Figure~\ref{fig:sigma} show that both the peak position
(in cm$^{-1}$) and width decrease with increasing temperature, but they are not
sensitive to CO$_2$ concentration. This isolates the effect of temperature from
the effect of concentration, and thus provides an easy tool for the
determination of the thermal history. Next we try to fit the $^{13}$CO$_2$
spectra in \citet{Boogert2000} based on our laboratory measurements. We
visually examine the observed spectra and separate them into two groups: group
1 without a significant narrow blue peak at 2283 cm$^{-1}$, and group 2 with
it. Spectra in group 1 are fit with a single Gaussian function, while spectra
in group 2 are fit with one Gaussian function for disordered CO$_2$ and one
Lorentzian function for crystalline CO$_2$, as shown in Figure
\ref{fig:1comp} and \ref{fig:2comp}. The best fit peak position, full width at
half maximum ($FWHM$) as well as the calculated temperature based on the Gaussian
component using Equation \ref{eq:sigma} are also shown for each spectrum. For
group 2, the magnitude of both components are also shown.

The temperatures quoted in Figure~\ref{fig:1comp} and \ref{fig:2comp} are based
on the laboratory time scale. In a typical astrophysically relevant time
scale, the warming up of the ice is over a much longer time, and a lower
temperature is expected to yield the same structural changes in the ice
mixture. To translate the ice's temperature in the laboratory time scale
$T_{\rm lab}$ to the temperature in astronomical conditions $T_{\rm ast}$, an
Arrhenius-type expression can be used to describe the rate of segregation.
\begin{equation}
k=\nu \exp(-E_{\rm seg}/T)
\end{equation}
where $\nu$ and
$E_{\rm seg}$ are the prefactor and the energy barrier (in unit of degree Kelvin) for
CO$_2$ segregation, respectively. The time scale for segregation to happen can
be approximated by $t\sim k^{-1}$. The temperature at which segregation happens
most efficiently is:
\begin{equation}
	T=\frac{E_{\rm seg}}{ln(\nu t)}
\end{equation}
In laboratory experiments, the time scale $t_{\rm lab}$ is in the order of
seconds. Under astronomical conditions, the free fall time between distances
that have temperature of 70--90 K for a low mass star is 75 yrs
\citep{Pontoppidan2008}. Rotation will slow down this process, but
10$^2$--10$^3$ yrs for the temperature range where segregation takes place is
a reasonable estimate. The lifetime of hot cores around massive YSOs is 30,000
yrs \citep{Charnley1992}. These hot cores have temperatures exceeding 100 K, but
there is a gradient outside of that where ices have not yet evaporated and are
heated during that time. We adopt the range of 10$^2$--10$^5$ yrs for the
segregation time scale under astronomical conditions. The conversion
factor from laboratory time scale to astronomical time scale is:
\begin{equation}
	\frac{T_{\rm ast}}{T_{\rm lab}} = \frac{ln(\nu
t_{\rm lab})}{ln(\nu t_{\rm ast})}
\end{equation}
The prefactor $\nu$ is not well characterized. \citet{Oberg2009} did isothermal
experiments of 1:2 CO$_2$:H$_2$O mixtures and reported a prefactor of
$2\times10^{5\pm 1}$ s$^{-1}$ for segregation. This prefactor likely reflects
a combined effects of CO$_2$ diffusion on ASW and the collapse of pores in ASW.
Using this prefactor, the conversion factor can be calculated to be in the
0.3--0.4 range. If we assume that the segregation is dominated by the diffusion
of CO$_2$ on the pore surface of ASW and take the laboratory determined
prefactors for volatile molecules such as CO, N$_2$, CH$_4$, which are mostly
in the $10^{8\pm 1}$ range \citep{He2018diffusion}, then the conversion factor
is in the 0.4--0.5 range. In summary, to convert the temperature in the
laboratory time scale to that of the warming up of a interstellar cloud,
a factor 0.4$\pm$0.1 should be considered. With a correction for time scale
taken into account, the dependence of segregation on both concentration and
temperature shown in Figure~\ref{fig:linear_doc} should be useful in models of
YSO envelopes.

\subsection{Comparison with Astronomical Spectra}
In the sightline of W33A, the $^{13}$CO$_2$ absorption peak is centered at
2277 cm$^{-1}$, which corresponds to a temperature of 105 K in the laboratory
time scale. But if we plug $T=105$ K into Eq~\ref{eq:sigma}, the calculated
$FWHM=6.3$ cm$^{-1}$ is much smaller than that for the observed one 9.0
cm$^{-1}$. Note that this is likely not due to contamination by CO mixed in the
ices, because W33A has less contribution from CO (10\%) compared to other YSOs,
e.g., NGC 7538 IRS9 (20\%; \citet{Pontoppidan2008}). This inconsistency is
likely due to the non-uniform temperature along the sightline. For this
reason, we use the peak position in Eq~\ref{eq:mu} instead of the $FWHM$ in
Eq~\ref{eq:sigma} to determine the temperature. In general, the best fit
temperature in Figure \ref{fig:1comp} should be understood as the average
temperature along the sightline.

So far, we used CO$_2$ in water ice to obtain the temperature of the ice. Now
we put the temperature traced by $^{13}$CO$_2$ in the context of the
observed sightlines. Of the targets with a single ``disordered'' $^{13}$CO$_2$
component (Figure~\ref{fig:1comp}), the peak of R CrA IRS2 (TS13) has the largest
wavenumber, and thus the lowest temperature. The space temperature of
(0.4$\pm$0.1)$\times$(23$\pm$12) K is well below that for CO sublimation, and
indeed this sightline harbors an exceptionally large apolar CO component
\citep{Vandenbussche1999}. The two other high quality spectra of the low mass
YSO HH 100 IR and the massive YSO W33A show $^{13}$CO$_2$ bands peaking at
smaller wavenumbers, corresponding to space temperatures of
(0.4$\pm$0.1)$\times$(55$\pm$9) K and (0.4$\pm$0.1)$\times$(105$\pm$7) K,
respectively. Indeed, the CO profile of W33A is dominated by polar CO ices
\citep{Pontoppidan2003}, indicating that the volatile apolar CO ices have
sublimated, and a significant abundance of warm CO gas is detected in this
sightline \citep{Mitchell1990}. The relatively high $^{13}$CO$_2$ temperature
for HH 100 seems puzzling considering the large abundance of apolar CO
\citep{Pontoppidan2003} . This likely reflects a large temperature gradient in
the HH 100 YSO envelope, with colder apolar CO dominating the ices at larger
radii. One should also consider the possibility that the CO$_2$:H$_2$O ices are
formed earlier in the cloud history in less shielded conditions on warmer
grains than the volatile CO. Then the $^{13}$CO$_2$ profile reflects the
formation temperature. This can be tested by observations towards background
stars tracing quiescent clouds. Unfortunately, the ISO/SWS spectrum of Elias
16, a background star of the Taurus Molecular Cloud, is of low quality, but
will be vastly improved when observed with JWST in the near future. For
a discussion regarding the targets with segregated crystalline $^{13}$CO$_2$
(Figure 12) we refer to \citet{Boogert2000} and \citet{VanderTak2000}, who show
that the degree of segregation correlates with the dust temperature as the YSO
envelope becomes less massive and hotter over time scales of a few times 10,000
yrs.

In the second group (Figure~\ref{fig:2comp}), we used one broad Gaussian and
one narrow Lorentzian to fit the spectra. The temperatures are derived from the
broad component using Eq~\ref{eq:mu}. \citet{Boogert2000} used a narrow
Gaussian instead of a Lorentzian to fit the 2283 cm$^{-1}$ feature. They found
that the narrow feature is present only toward high mass protostars, even though
not all high mass protostars have the narrow feature (see \citet{Boogert2000}
for a detailed discussion). In these four sightlines, the temperatures are all
relatively high, in agreement with the previous proposition that segregation
indicates thermal processing. Because of more free parameters being used in the
fitting, the uncertainty in temperature is much larger than in group 1.
Furthermore, the signal-to-noise ratios of the spectra from S140:IRS1 and
W3:IRS5 are not good enough for an accurate determination of the peak position of
the disordered component. To better constrain the temperature would require better
signal-to-noise ratio of the spectra. The James Webb Space Telescope (JWST) is
expected to cover these regions at orders of magnitude better sensitivity and
somewhat larger spectral resolution ($R=3,000$ versus 2,000), and more accurate
temperature determination and sorting of features between different classes of
objects will be possible. The bending mode profile at 15 $\mu$m can also be
used as a supplementary tool to further constrain the temperature.

Previously, \citet{Boogert2000} compared the observed spectra with the
laboratory measurement of \citet{Ehrenfreund1999}. They found a similar
red-shift and a narrowing of the $^{13}$CO$_2$ peak as the temperature of the
CO$_2$:H$_2$O mixture was increased. They also found that during heating
a 1:0.92:1 H$_2$O:CH$_3$OH:CO$_2$ mixture, the peak position and width changed,
but with a different temperature dependence than that of a CO$_2$:H$_2$O
mixture. Therefore, \citet{Boogert2000} concluded that while $^{13}$CO$_2$
traces segregation, it is not suitable for temperature determinations, because
the temperature effect could not be separated from composition effects. In their
experiment, the concentration of CH$_3$OH is much higher than what is typically
observed. CH$_3$OH is formed mostly on dust grains by the consecutive
hydrogenation of CO after the heavy CO freeze-out. This formation mechanism is
corroborated by both laboratory measurements \citep{Hama2013} and observations
\citep{Boogert2011}, which show that CH$_3$OH is mostly found in high
extinction regions. Conversely, CO$_2$ is mostly found in a water-rich
environment, consistent with the scenario that CO$_2$ is formed together with
water. Although most recent laboratory experiments found that CH$_3$OH can also
be formed before the heavy freeze-out of CO by the reaction between CH$_3$ and
OH \citep{Qasim2018}, the question still remains of how much CH$_3$OH can be
formed this way. Based on the current state of knowledge, it is safe to assume
that CO$_2$ mostly interacts with water instead with CH$_3$OH in the ice, and
therefore it is justifiable to ignore the effect of CH$_3$OH on the
$^{13}$CO$_2$ absorption profile.

\subsection{CO:CO$_2$ Ices}
This assumption seems less applicable to CO mixed in the ices. Previous
analysis of the CO$_2$ bending mode absorption profile at 15 $\mu$m and the
blue component of CO absorption profile at 4.7 $\mu$m reveals that 10-30\% of
the CO$_2$ molecules are mixed with CO \citep{Pontoppidan2008,
Pontoppidan2003}. This group proposed that ``pure'' crystalline CO$_2$ can be
formed either by thermally-induced CO$_2$ segregation from a CO$_2$:H$_2$O
mixture, or by CO desorption from a CO:CO$_2$ mixture. The mechanism of
segregation is closely related to the mechanism of CO$_2$ formation. So far
there are mainly two categories of mechanisms proposed to explain the formation
of CO$_2$ molecules on grains. The first category involves pure thermal
reactions among CO, O, H, and OH. Although several questions still remains, such
as the relative contribution of CO+O \citep{Roser2001} and CO+OH
\citep{Zins2011}, and whether the reaction involves HOCO \citep{Ioppolo2011},
it is clear that water is formed on grains at the same time as CO$_2$ via
reactions with hydrogen: O+H$\rightarrow$OH, OH+H$\rightarrow$H$_2$O. The experimental results and
the temperature tracing method proposed in this study mostly apply to the
CO$_2$ that is formed together with water by thermal processes. The second
category of CO$_2$ formation mechanism involves energetic processing
of the pure CO in the top layers of the ice mantle. Laboratory experiments have
shown that CO$_2$ can be formed by the bombardment of analogues of cosmic rays
with CO ice \citep{Gerakines2001,Loeffler2005,Jamieson2006}. However, other
molecules that should have also been produced in the energetic processing of
CO, such as C$_3$O$_2$ \citep{Jamieson2006}, were not observed in the
same sightline as CO$_2$ \citep{Pontoppidan2008}. The reason why a fraction
of CO$_2$ is in CO-rich environment is still puzzling. If cosmic ray
bombardment is important for CO$_2$ formation, the compaction of the ASW and
the segregation of CO$_2$ from CO$_2$:H$_2$O mixtures should also be affected by
cosmic rays. It would be less relevant to characterize the ice mantle by
temperature than to use the fluence of cosmic rays. In any case, the use of
CO$_2$ segregation or $^{13}$CO$_2$ to trace the temperature history of the ice
mantle is only valid under the assumption that cosmic ray irradiation is not
the dominant mechanism for CO$_2$ formation.
\section{Summary}
We made measurements in the laboratory of infrared absorption features of CO$_2$:H$_2$O ices with different mixing ratios, at different temperatures, and subjected to thermal cycles in order to elucidate the thermal history of ices observed towards YSOs with ISO-SWS \citep{Boogert2000}. We found that at a CO$_2$:H$_2$O concentration below 23\%, there is no formation of pure crystalline CO$_2$. Thus, looking at pure crystalline CO$_2$ alone is not a good proxy for establishing the thermal history of ices. We found that the $\nu_3$ feature of $^{13}$CO$_2$ does not suffer from this threshold concentration dependence, and its peak position and linewidth, together with the pure crystalline feature of $^{12}$CO$_2$, can be used to infer the temperature history of ices near YSOs. Data such as the ones presented here will help to characterize the segregation status and thermal history of ices in upcoming JWST observations with an extended range in the IR spectrum, and improved sensitivity and spectral resolution.

\section*{Acknowledgements}
We thank Francis Toriello for technical assistance.
This research was supported by NSF Astronomy \& Astrophysics Research Grant
\#1615897.

\appendix
\section{Deposition of gases}
CO$_2$ gas and water vapor were deposited onto the sample through background
deposition using two UHV precision leak valves activated by two stepper motors
controlled by a LabVIEW program. For the deposition of a single molecular
species, the program first measures the base pressure of the chamber, and then
calculates the target partial pressure based on a user set deposition rate. The
pressure readings from the hot cathode ion gauge is corrected for the gas
species in the LabVIEW program. A PID control loop is used to maintain the
pressure at the target value. In the deposition of CO$_2$, it takes
about 20 seconds for the pressure to stabilize at the set value. The ice
thickness during deposition is calculated by the program in real time. When the
thickness reaches the setpoint, the leak valve is closed quickly. Even after
the valve is closed, the residual gas in the chamber continues to being
deposited on the sample, until it is pumped out. We correct for the additional
amount deposited from the residual gas by closing the valve slight before the
target thickness is reached. The exact offset thickness is calculated from the
deposition pressure and the pumping speed. After this correction, in the
deposition of CO$_2$, the {\em{relative}} uncertainty of thickness measured by
the integration of pressure over time is usually less than 0.1\%. For water
deposition, the uncertainty is larger (1\%) because of the difficulty in
maintaining a stable water inlet pressure in the gas manifold.

In CO$_2$:H$_2$O co-depositions, because the ion gauge can only measure the
total pressure but not the partial pressure of each gas, we start with the
deposition of one gas. CO$_2$ is deposited first because it is easier to
control its deposition. Within 20 seconds of introducing CO$_2$, the deposition
rate is already stable. We tested the stability of pressure by fixing the valve
position after 20 seconds, and found that the pressure does not change over
time. The same is not true for water because of the instability of inlet
pressure. After finding out the stable valve position for
CO$_2$, we stop the PID loop for the CO$_2$ valve and fix the valve position.
We then use a PID loop for the water leak valve to obtain a stable pressure
during co-deposition. When the set time is reached, both leak valves are closed
immediately. For the co-depositions in this study, the deposition is over 25
minutes, and therefore the uncertainty in CO$_2$ amount is about 20 seconds
over 25 minutes, which is about 1\%.

The impingement rate ($IPR$), which is the number of molecules deposited per
unit surface area per unit time, is calculated as follows:
\begin{equation}
IPR=\frac{P}{\sqrt{2\pi m k_{\rm B} T}}
\end{equation}
where $P$ is the chamber pressure after correction for the ion gauge
gas specific ionization cross-section, $m$ is the mass of gas molecule, and $T$ is the gas temperature (assumed to be room temperature),
$k_B$ is the Boltzmann constant. It is assumed that the sticking of both CO$_2$
and H$_2$O are unity at 10 K \citep{He2016a}, and the pressure in the vacuum
chamber is uniform. This is a fair assumption, because the leak valves opening
does not face the sample or cold head directly. The impingement rate can be
converted to the unit of monolayer per second (ML/s) by assuming 1 ML
= 10$^{15}$ cm$^{-1}$. The absolute uncertainty of deposition is mostly due to
the uncertainty in pressure measurement, and can be as high as 30\%, as this is
the accuracy of a typical hot cathode ion gauge. In the experiments, the
uncertainty in mixing ratio of the CO$_2$:H$_2$O mixtures is governed by the
relative uncertainty, while that of the total thickness of the mixture is
governed by the absolute uncertainty.

\end{document}